\documentclass[12pt]{iopart}
\usepackage{iopams}
\usepackage{graphicx}
\usepackage{tikz}
\usetikzlibrary{positioning}
\usepackage[hidelinks]{hyperref}

\newcommand{\sech}{\mathrm{sech}}

\makeatletter

\def\endequation{\eqno\hbox{\@eqnnum}$$\@ignoretrue}
\makeatother
\begin{document}

\title[Thermodynamic geometry of three coupled Ising spins]{Thermodynamic geometry of three coupled Ising spins: exact curvature, frustration and bond anisotropy}

\author{Zahra Ebadi$^{1}$, Nayyere Einali Saghavaz$^{1}$, Omid Yahyayi Monem$^{1}$ and Hosein Mohammadzadeh$^{1,\ast}$}

\address{$^{1}$Department of Physics, University of Mohaghegh Ardabili, P.O. Box 179, Ardabil, Iran}

\eads{$^{\ast}$\mailto{mohammadzadeh@uma.ac.ir}}

\begin{abstract}
Thermodynamic curvature is widely used as a diagnostic of interactions and phase transitions, yet its interpretation is usually tested only in the thermodynamic limit. Small spin clusters are physically relevant in their own right, since a macroscopic sample of independent molecular trimers or a few-qubit quantum simulator is governed exactly by the few-spin partition function. They also provide the only setting in which the geometry can be obtained in closed form together with its full dependence on the interaction topology. We study the Fisher--Rao (Ruppeiner) geometry of three Ising spins with arbitrary pairwise couplings in a longitudinal field. This cluster interpolates continuously between an open chain and a geometrically frustrated triangle. For the symmetric triangle we obtain the scalar curvature in closed form. It vanishes on the noninteracting line and on a line of fixed field-to-temperature ratio that does not depend on the coupling, which coincides with the inflection point of the single-spin susceptibility. The near-zero behaviour is therefore exactly linear and reflects analyticity rather than critical scaling. In the frustrated antiferromagnetic regime the curvature remains finite and negative down to zero temperature. A general low-temperature formula relates this plateau to the change in magnetization content between the degenerate ground manifold and its first excitation. Any bond anisotropy destroys the plateau below a crossover temperature that depends only logarithmically on the anisotropy, so the curvature is an extremely sensitive detector of exact ground-state degeneracy. On the ferromagnetic side the zero-temperature curvature is universal for all couplings, whereas additional zero lines in anisotropic clusters track field-induced crossings of excited levels.
\end{abstract}

\vspace{2pc}
\noindent Keywords: Ising model, frustrated spin systems, classical spin models, thermodynamic geometry

\maketitle

\section{Introduction}\label{sec:intro}

The Ising model, proposed by Lenz \cite{lenz1920} and solved in one dimension by Ising \cite{ising1925beitrag}, is the paradigmatic model of cooperative behaviour in statistical mechanics \cite{brush1967history,baxter2016exactly}. With short-range interactions it has no phase transition at finite temperature in one dimension \cite{ising1925beitrag,ruelle1969statistical}. The energy--entropy argument of Landau and Peierls \cite{peierls1936ising} explains why discrete-symmetry order requires $d\geq 2$, as confirmed by Onsager's solution \cite{onsager1944crystal}; the Mermin--Wagner theorem \cite{mermin1966absence} is the analogous statement for continuous symmetries.

Thermodynamic geometry endows the space of equilibrium states with a Riemannian metric whose line element measures the distinguishability of neighbouring states through their fluctuations \cite{Weinhold1975,Ruppeiner1979,Salamon1984ThermoLength,Ruppeiner1995,mrugala1990,crooks2007}. In the statistical formulation the metric is the Fisher information of the Gibbs distribution \cite{amari2000,brody1995,Janyszek1990}. Its scalar curvature $R$ is commonly given three interpretations \cite{Ruppeiner1995,ruppeiner2010ajp,Janyszek1990QuantumGases}. First, $R=0$ for noninteracting systems, and the sign of $R$ distinguishes effectively attractive from repulsive statistical interactions, as for ideal Bose and Fermi gases. Second, $|R|$ measures the volume of correlated fluctuations, $|R|\sim\xi^d$ near criticality. Third, $R$ diverges at phase transitions. The approach has been applied to quantum gases and generalized statistics \cite{Janyszek1990QuantumGases,mirza2010thermodynamic,eglinton2023thermodynamic,adli2019nonperturbative,esmaili2024thermodynamic,seifi2025intrinsic,seifi2025mittag,ardabili2025haldane,yahyayi2026thermodynamic,mohammadzadeh2026thermodynamic}, to fluids and the Widom line \cite{ruppeiner2012thermodynamic,ruppeiner2012widom,Quevedo2022GTDvdW}, to black holes \cite{BabaeiAghbolagh2022TGComplexity}, and to classical spin models, including the one-dimensional Ising chain \cite{janyszek1989riemannian,janke2004}, the Bethe lattice \cite{dolan1998} and the kagome lattice \cite{mirza2013thermodynamic}. Its information-theoretic counterpart is now a standard tool in the study of quantum phase transitions \cite{zanardi2007,sachdev2011}.

A natural objection to studying only three spins is that such a system has no thermodynamics in the usual sense. We have three reasons for considering it nonetheless. First, the canonical description of a small system in contact with a heat bath is exact and needs no thermodynamic limit; this is the basis of the thermodynamics of small systems \cite{hill1963} and of stochastic thermodynamics \cite{seifert2012}. The Fisher--Rao metric is defined for any Gibbs distribution, whatever the number of degrees of freedom. Second, three-spin clusters are realized physically. In molecular magnets, a macroscopic crystal of magnetically isolated trimers has a total partition function $\mathcal Z_{\rm tot}=\mathcal Z_3^{N_{\rm mol}}$ \cite{kahn1993,gatteschi2006}. Its thermodynamics is therefore genuinely macroscopic, yet it is fixed exactly by the three-spin partition function, and its curvature differs from the three-spin curvature only by the factor $1/N_{\rm mol}$. Frustrated three-spin Ising Hamiltonians have also been engineered directly in trapped-ion quantum simulators \cite{kim2010}, and few-spin systems serve as working media of thermal machines \cite{henrich2007driven,ahadpour2021coupled,abd2023comparative}. Third, three is the smallest size at which the topology of the interaction graph matters. A triangle can be frustrated \cite{wannier1950,toulouse1977,moessner2006} whereas an open chain cannot, and the exact solution allows us to test how frustration, degeneracy and bond anisotropy appear in the geometry without the uncertainties of approximate or numerical treatments of large systems.

For finite systems the second and third interpretations of $R$ lose their direct meaning, because $R$ is an analytic function of the parameters. Brody and Ritz showed that finite Ising chains nevertheless divide their parameter space into domains of positive and negative curvature, which merge as $N\to\infty$ \cite{brody1999geometric}. In a previous work we analysed two coupled spins with several Ising Hamiltonians \cite{saghavaz2025thermodynamic}. Here we take the next step to three spins with three pairwise couplings $J$, $uJ$ and $aJ$. This cluster interpolates continuously between an open chain ($a=0$) and a symmetric triangle ($a=u=1$), which is frustrated for antiferromagnetic coupling.

Our main results are as follows. For the symmetric triangle the curvature has a compact closed form that factorizes into a coupling part and a field part. Its zero set consists of the line $x=0$ and a line of fixed $y$ that is independent of temperature and coupling, which we identify with the inflection point of the single-spin susceptibility. An apparent power law near this line therefore has exponent one exactly. A general low-temperature formula shows that the finite negative curvature of the frustrated triangle at zero temperature reflects the magnetization content of the lowest excitation above the degenerate frustrated manifold. Any bond anisotropy destroys this plateau below a crossover temperature inversely proportional to the logarithm of the anisotropy, so that $R$ is exponentially sensitive to exact degeneracy. On the ferromagnetic side the zero-temperature curvature is universal for all couplings, whereas additional zero lines in anisotropic clusters follow field-induced crossings between excited levels.

The paper is organized as follows. Section~\ref{sec:model} introduces the model and its partition function, and section~\ref{sec:geometry} summarizes the geometric framework and fixes the sign convention. Section~\ref{sec:triangle} presents the exact solution for the symmetric triangle, section~\ref{sec:aniso} treats weak bond anisotropy, and section~\ref{sec:chain} follows the interpolation from the open chain to the triangle. Section~\ref{sec:discussion} discusses the results and section~\ref{sec:conclusion} concludes. Technical details are collected in the appendices.

\section{Model and partition function}\label{sec:model}

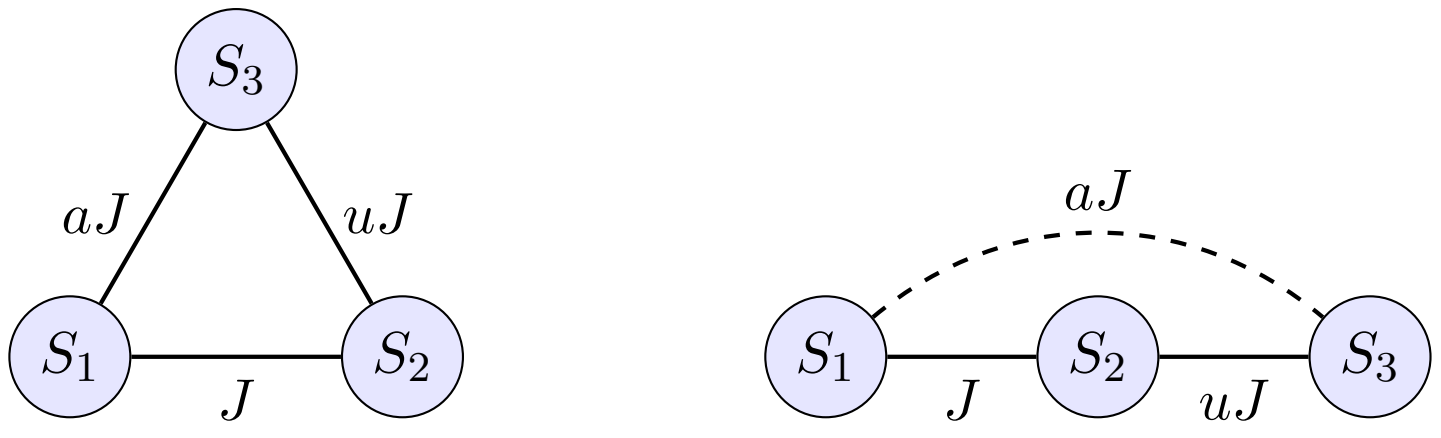
\begin{figure}[t]
\centering
\begin{tikzpicture}[scale=1.0, every node/.style={font=\small}]
\begin{scope}[xshift=0cm]
\node[circle,draw,fill=blue!10,minimum size=8mm,inner sep=0pt] (S1) at (0,0) {$S_1$};
\node[circle,draw,fill=blue!10,minimum size=8mm,inner sep=0pt] (S2) at (2.2,0) {$S_2$};
\node[circle,draw,fill=blue!10,minimum size=8mm,inner sep=0pt] (S3) at (1.1,1.9) {$S_3$};
\draw[thick] (S1)--node[below]{$J$}(S2);
\draw[thick] (S2)--node[right]{$uJ$}(S3);
\draw[thick] (S3)--node[left]{$aJ$}(S1);
\end{scope}
\begin{scope}[xshift=5cm]
\node[circle,draw,fill=blue!10,minimum size=8mm,inner sep=0pt] (L1) at (0,0) {$S_1$};
\node[circle,draw,fill=blue!10,minimum size=8mm,inner sep=0pt] (L2) at (1.8,0) {$S_2$};
\node[circle,draw,fill=blue!10,minimum size=8mm,inner sep=0pt] (L3) at (3.6,0) {$S_3$};
\draw[thick] (L1)--node[below]{$J$}(L2);
\draw[thick] (L2)--node[below]{$uJ$}(L3);
\draw[thick,dashed] (L1) to[bend left=40] node[above]{$aJ$} (L3);
\end{scope}
\end{tikzpicture}
\caption{The three-spin Ising cluster of equation~(\ref{eq:H}): triangle (left) and chain with optional 1--3 bond (right, dashed).}
\label{fig:schematic}
\end{figure}

We consider three Ising spins $S_i=\pm1$ with Hamiltonian
\begin{equation}
H=-J\left(S_1S_2+u\,S_2S_3+a\,S_1S_3\right)-h\left(S_1+S_2+S_3\right),
\label{eq:H}
\end{equation}
where $J$ sets the coupling scale ($J>0$ ferromagnetic, $J<0$ antiferromagnetic), $h$ is a longitudinal field, and the dimensionless ratios $u$ and $a$ fix the strengths of the 2-3 and 1--3 bonds relative to the 1--2 bond. The case $a=u=1$ is the symmetric triangle and $a=0$ is the open three-site chain; intermediate values interpolate between the two topologies, as sketched in figure~\ref{fig:schematic}.

In the triangle every spin interacts with the other two. In the chain only nearest neighbours interact when $a=0$, and the dashed bond of strength $0<a<1$ connects the two topologies. It is convenient to write
\begin{equation}
H=-JB-hM,
\end{equation}
in terms of the bond variable and the magnetization,
\begin{equation}
B=S_1S_2+uS_2S_3+aS_1S_3,\qquad M=S_1+S_2+S_3 .
\end{equation}
The eight configurations, grouped into pairs related by global spin reversal, are listed in table~\ref{tab:spectrum}; each row also represents its reversed partner, which has the same $B$ and opposite $M$.

\begin{table}[t]
\caption{Bond sum $B$, magnetization $M$ and energy $E$ of the three-spin configurations.}
\label{tab:spectrum}
\centering
\begin{tabular}{cccccc}
\br
$S_1$ & $S_2$ & $S_3$ & $B$ & $M$ & $E$\\
\mr
$+$ & $+$ & $+$ & $1+u+a$ & $3$ & $-J(1+u+a)-3h$\\
$+$ & $+$ & $-$ & $1-u-a$ & $1$ & $-J(1-u-a)-h$\\
$+$ & $-$ & $+$ & $-1-u+a$ & $1$ & $J(1+u-a)-h$\\
$-$ & $+$ & $+$ & $-1+u-a$ & $1$ & $J(1-u+a)-h$\\
\br
\end{tabular}
\end{table}

With the dimensionless intensive variables
\begin{equation}
x=\beta J,\qquad y=\beta h,\qquad \beta=\frac{1}{k_BT},
\end{equation}
the partition function $\mathcal Z=\sum e^{xB+yM}$ reads
\begin{equation}
\begin{array}{rl}
\mathcal{Z}(x,y)=&2e^{x(1+u+a)}\cosh 3y+2e^{x(1-u-a)}\cosh y
+\,4e^{-x}\cosh\left[x(u-a)\right]\cosh y ,
\end{array}
\label{eq:Z}
\end{equation}
which we checked against direct enumeration of the eight states. Since $\mathcal Z$ is even in $y$, all quantities below are invariant under $y\to-y$, and we restrict ourselves to $y\geq0$. For the symmetric triangle, the identity
\begin{equation}
\cosh 3y=\cosh y\,(2\cosh 2y-1)
\end{equation}
simplifies equation~(\ref{eq:Z}) to
\begin{equation}
\mathcal{Z}_\triangle=2e^{-x}\cosh y\left[e^{4x}\left(2\cosh 2y-1\right)+3\right].
\label{eq:Ztri}
\end{equation}

\section{Thermodynamic geometry}\label{sec:geometry}

The Gibbs distribution
\begin{equation}
p(\sigma)=\frac{e^{xB(\sigma)+yM(\sigma)}}{\mathcal Z}
\end{equation}
is an exponential family with natural parameters $\theta=(x,y)$ and sufficient statistics $(B,M)$. Its Fisher--Rao metric \cite{amari2000,brody1995,Janyszek1990} is the Hessian of $\ln\mathcal Z$,
\begin{equation}
g_{ij}=\frac{\partial^2\ln\mathcal Z}{\partial\theta^i\partial\theta^j},
\label{eq:metric}
\end{equation}
which is the covariance matrix of $B$ and $M$,
\begin{equation}
g_{xx}=\langle\Delta B^2\rangle,\qquad g_{xy}=\langle\Delta B\,\Delta M\rangle,\qquad g_{yy}=\langle\Delta M^2\rangle .
\end{equation}
This metric is the Legendre-dual form of Ruppeiner's entropy-Hessian metric \cite{Ruppeiner1979,Ruppeiner1995} and is conformally related to Weinhold's energy metric \cite{Weinhold1975,Salamon1984ThermoLength}. It is positive definite for every finite $(x,y)$, so the geometry is regular everywhere.

For a two-dimensional Hessian metric the scalar curvature takes the determinant form \cite{Janyszek1990QuantumGases,brody1999geometric,Ruppeiner1995}
\begin{equation}
R=-\frac{1}{2G^2}
\left|
\begin{array}{ccc}
g_{xx} & g_{xy} & g_{yy}\\
g_{xx,x} & g_{xy,x} & g_{yy,x}\\
g_{xx,y} & g_{xy,y} & g_{yy,y}
\end{array}\right|,\qquad G=\det g,
\label{eq:R}
\end{equation}
where $g_{ij,k}=\partial_kg_{ij}$. Since $g$ is a Hessian, the derivatives $g_{ij,k}$ are third-order cumulants. Introducing the joint cumulants of $B$ and $M$,
\begin{equation}
\kappa_{mn}=\frac{\partial^{m+n}\ln\mathcal Z}{\partial x^m\,\partial y^n},
\end{equation}
the curvature is fixed entirely by second- and third-order fluctuations,
\begin{equation}
R=-\frac{1}{2\left(\kappa_{20}\kappa_{02}-\kappa_{11}^2\right)^2}
\left|
\begin{array}{ccc}
\kappa_{20} & \kappa_{11} & \kappa_{02}\\
\kappa_{30} & \kappa_{21} & \kappa_{12}\\
\kappa_{21} & \kappa_{12} & \kappa_{03}
\end{array}\right|.
\label{eq:Rcumulant}
\end{equation}

The sign of $R$ depends on convention, so we fix it explicitly. We evaluated equation~(\ref{eq:R}) for the ideal quantum gases, for which
\begin{equation}
\ln\mathcal Z=\beta^{-3/2}f_{5/2}\left(e^{-\gamma}\right),\qquad \gamma=-\beta\mu,
\end{equation}
in the coordinates $(\beta,\gamma)$. At $\beta=\gamma=1$ we find $R=+0.250$ for bosons and $R=-0.205$ for fermions, in agreement with references \cite{Janyszek1990QuantumGases,mirza2010thermodynamic}. Throughout this paper, positive (negative) curvature therefore signals effectively attractive (repulsive) statistical correlations, in the sense in which Bose (Fermi) statistics are attractive (repulsive). For a finite cluster this is a statement about the correlations between the fluctuations of $B$ and $M$, not about a microscopic force, and we use it only in that sense.

\section{Symmetric triangle: exact curvature}\label{sec:triangle}

\subsection{Closed form}

For $a=u=1$ the linear prefactor $e^{-x}$ in equation~(\ref{eq:Ztri}) does not contribute to the metric. Evaluating equation~(\ref{eq:R}) (\ref{app:closed}) gives the closed result
\begin{equation}
R_\triangle(x,y)=-\frac{8\,e^{4x}\sinh(4x)\,\cosh^2y\,\left(\cosh 2y-2\right)}{\left(3e^{4x}+2\cosh 2y-1\right)^2}.
\label{eq:Rtri}
\end{equation}
The denominator is strictly positive, so the sign of the curvature is fixed by the numerator,
\begin{equation}
\mathrm{sgn}\,R_\triangle=-\,\mathrm{sgn}(x)\;\mathrm{sgn}(y-y_c),
\label{eq:sign}
\end{equation}
where $y_c$ is defined by
\begin{equation}
\cosh 2y_c=2,\quad y_c=\frac{1}{2}\ln\left(2+\sqrt3\right)=0.65848\ldots,\quad \tanh^2y_c=\frac{1}{3}.
\label{eq:yc}
\end{equation}

\begin{figure}[t]
\centering
\includegraphics[width=0.55\textwidth]{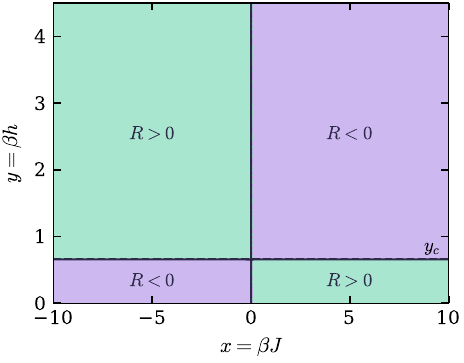}
\caption{Sign of $R$ versus $x=\beta J$ and $y=\beta h$ for the symmetric triangle, $a=u=1$.}
\label{fig:triangle}
\end{figure}

Figure~\ref{fig:triangle} shows the resulting sign map. The green regions correspond to $R>0$ and the violet regions to $R<0$. The parameter plane splits into four quadrants bounded by two straight lines, $x=0$ and $y=y_c$; the latter is shown as a dashed line. The pattern does not depend on the plotted range. This clarifies what numerical maps alone suggest: the boundary at $y_c$ is exactly horizontal and independent of $J$ and $T$. It is therefore a property of the geometry and not a crossover temperature.

\subsection{Meaning of the two zero lines}

At $x=0$ the spins are independent, and for the triangle the curvature vanishes there, as expected for a noninteracting system \cite{Ruppeiner1995,janke2004}. Close to this line,
\begin{equation}
R_\triangle\simeq 2x\left(1-3\tanh^2y\right),
\label{eq:smallx}
\end{equation}
so for $y<y_c$ the sign of $R$ follows the sign of the coupling. Ferromagnetic coupling gives $R>0$, an attractive, boson-like correlation consistent with the tendency to align. Antiferromagnetic coupling gives $R<0$, a repulsive, fermion-like correlation. As shown in section~\ref{sec:chain}, the vanishing of $R$ at $x=0$ is not generic: for anisotropic clusters $R$ is nonzero at $x=0$ although the spins are free there. The Fisher geometry depends on which statistic is conjugate to the coordinate $x$, here $B$. For the triangle $B$ is permutation symmetric, and the cancellation is exact.

To interpret the line $y=y_c$ we rewrite the field factor as
\begin{equation}
\cosh 2y-2=-\cosh^2y\left(1-3\tanh^2y\right).
\end{equation}
For a single free spin, $\ln(2\cosh y)$ generates the magnetization cumulants. The second cumulant is the susceptibility, and the fourth is its second derivative:
\begin{equation}
\kappa_2^{(1)}(y)=\sech^2y,\qquad
\kappa_4^{(1)}(y)=\frac{d^2\kappa_2^{(1)}}{dy^2}=-2\,\sech^2y\left(1-3\tanh^2y\right).
\end{equation}
Equation~(\ref{eq:Rtri}) can therefore be written as
\begin{equation}
R_\triangle=-\frac{4\,e^{4x}\sinh(4x)\cosh^6y}{\left(3e^{4x}+2\cosh 2y-1\right)^2}\;\kappa_4^{(1)}(y).
\label{eq:kurt}
\end{equation}
The curvature changes sign exactly where the single-spin magnetization distribution changes from platykurtic ($\kappa_4^{(1)}<0$, a nearly symmetric two-point distribution at weak field) to leptokurtic ($\kappa_4^{(1)}>0$, a strongly polarized spin). Equivalently, $y_c$ is the inflection point of the single-spin susceptibility, where linear response gives way to saturation. Equation~(\ref{eq:sign}) then has a transparent reading. For $y<y_c$ the sign of $R$ is that of the coupling, whereas for $y>y_c$ the field reverses the sign of the effective statistical interaction. The geometry thus separates an interaction-dominated from a field-dominated regime. Because the field and coupling dependences factorize in the symmetric triangle, the separatrix is set by single-spin physics alone.

\subsection{Behaviour near the zero line and asymptotics}

\begin{figure}[t]
\centering
\includegraphics[width=0.9\textwidth]{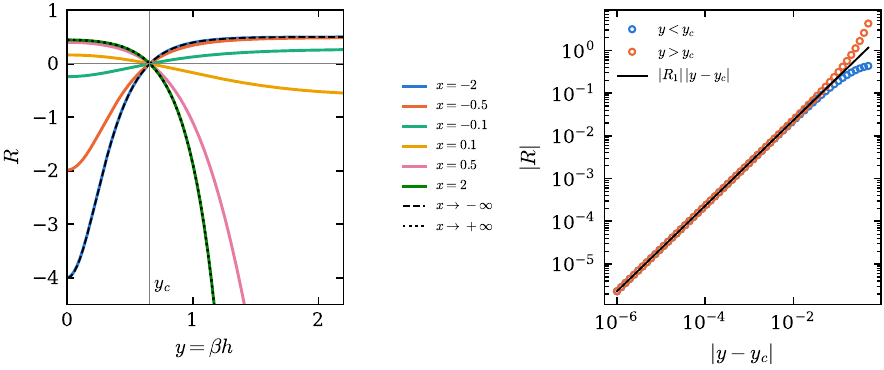}
\caption{Left: $R$ versus $y=\beta h$ for the symmetric triangle at several $x=\beta J$, with the limits $x\to\pm\infty$. Right: $|R|$ versus $|y-y_c|$ at $x=2$.}
\label{fig:profiles}
\end{figure}

Expanding equation~(\ref{eq:Rtri}) about $y_c$, using $\sinh 2y_c=\sqrt3$ and $\cosh^2y_c=3/2$, gives
\begin{equation}
R_\triangle\simeq-\frac{4}{\sqrt3}\tanh(2x)\,(y-y_c)+O\left((y-y_c)^2\right).
\label{eq:linear}
\end{equation}
If the curvature near $y_c$ is written as
\begin{equation}
R\propto|y-y_c|^{\gamma},
\label{eq:gamma}
\end{equation}
then $\gamma=1$ exactly, on both sides of $y_c$ and for every $x\neq0$. This is shown in the right panel of figure~\ref{fig:profiles}, where $|R|$ is plotted against $|y-y_c|$ at $x=2$ together with the linear law~(\ref{eq:linear}); the deviation at $|y-y_c|\gtrsim0.05$ comes from the quadratic correction. Fits of equation~(\ref{eq:gamma}) over a finite window give $\gamma\approx1.02$ below and $\gamma\approx0.98$ above $y_c$, and these small departures from unity are due entirely to the quadratic term. More generally, a simple zero of an analytic function always gives $\gamma=1$, so such an exponent should not be read as evidence of critical-like scaling. Genuine critical scaling of the curvature,
\begin{equation}
R\sim\xi^d\sim|t|^{\alpha-2},
\end{equation}
requires a divergent correlation length \cite{Ruppeiner1995,dolan1998,janke2004}; for the kagome Ising model, which has $\alpha=0$, Mirza and Talaei found $R\propto|\beta-\beta_c|^{\alpha-1}$ \cite{mirza2013thermodynamic}. Such behaviour has no analogue in a three-spin system.

The coupling dependence is controlled by two limits, both finite at fixed $y$:
\begin{equation}
R_{+\infty}(y)\equiv\lim_{x\to+\infty}R_\triangle=-\frac{4}{9}\cosh^2y\left(\cosh 2y-2\right),
\label{eq:Rplus}
\end{equation}
\begin{equation}
R_{-\infty}(y)\equiv\lim_{x\to-\infty}R_\triangle=\frac{4\cosh^2y\left(\cosh 2y-2\right)}{\left(2\cosh 2y-1\right)^2}.
\label{eq:Rminus}
\end{equation}
At zero field,
\begin{equation}
R_\triangle(x,0)=\frac{8e^{4x}\sinh 4x}{\left(3e^{4x}+1\right)^2},
\label{eq:Rzero}
\end{equation}
which increases monotonically from $-4$ for strong antiferromagnetic coupling to $4/9$ for strong ferromagnetic coupling. The left panel of figure~\ref{fig:profiles} shows the profiles $R(y)$ for $x=\pm2,\pm0.5,\pm0.1$ together with the two limits (dashed and dotted black curves). All curves pass through zero at $y_c$, marked by the vertical line, and they approach the limits quickly, already for $|x|\gtrsim2$. On the antiferromagnetic side the curvature is bounded,
\begin{equation}
-4\leq R_{-\infty}(y)<\frac{1}{2},
\end{equation}
and it saturates at $1/2$ for large fields. On the ferromagnetic side the curvature grows without bound at large field,
\begin{equation}
R_{+\infty}(y)\simeq-\frac{e^{4y}}{18},\qquad y\gg1,
\end{equation}
in the regime where all fluctuations are exponentially frozen and the metric becomes nearly degenerate.

\subsection{Frustration and the low-temperature plateau}

For $J<0$ the three antiferromagnetic bonds cannot be satisfied simultaneously \cite{wannier1950,toulouse1977,moessner2006}. Every configuration other than the two fully polarized ones has two satisfied (antiparallel) bonds and one unsatisfied (parallel) bond, so that $B=-1$. The ground manifold at $h=0$ is therefore sixfold degenerate, with three states of $M=+1$ and three of $M=-1$, and has residual entropy $\ln 6$. The only excitation, at energy $4|J|$ above it, is the polarized pair with $M=\pm3$.

The finite limit~(\ref{eq:Rminus}) is the geometric fingerprint of this structure. \ref{app:lowT} shows that for $|x|\to\infty$ at fixed $y$, whenever $B$ is constant within the ground manifold, the curvature tends to
\begin{equation}
R\longrightarrow-\frac{\left[\ln\phi(y)\right]''}{2A''(y)},
\label{eq:lowT}
\end{equation}
where primes denote derivatives with respect to $y$. Denoting by $\mathcal{G}_0$ the set of ground-state configurations and by $\mathcal{G}_1$ the set of first-excited configurations, the magnetization content of these two manifolds is encoded in the restricted partition functions
\begin{equation}
\mathcal{Z}_n(y)=\sum_{\sigma\in\mathcal{G}_n}e^{yM(\sigma)},\qquad n=0,1,
\label{eq:Zn}
\end{equation}
and
\begin{equation}
A(y)=\ln\mathcal{Z}_0(y),\qquad \phi(y)=\frac{\mathcal{Z}_1(y)}{\mathcal{Z}_0(y)} .
\label{eq:Aphi}
\end{equation}
The energy gap and the multiplicities drop out. For the frustrated triangle,
\begin{equation}
A=\ln(6\cosh y),\qquad \phi\propto\frac{\cosh 3y}{\cosh y},
\end{equation}
and equation~(\ref{eq:lowT}) reproduces equation~(\ref{eq:Rminus}) in the equivalent form
\begin{equation}
R_{-\infty}(y)=-\frac{1}{2}\left(\frac{9\cosh^2y}{\cosh^23y}-1\right).
\end{equation}
Equation~(\ref{eq:lowT}) makes clear that the curvature does not measure the degeneracy itself. It measures how strongly the lowest excitation changes the magnetization content of the thermally accessible states. If the first excited manifold has the same magnetization distribution as the ground manifold, $\phi$ is constant and the leading curvature vanishes. This observation is the key to the next section.

\section{Weak bond anisotropy: lifting the frustration}\label{sec:aniso}

We now break the permutation symmetry by setting
\begin{equation}
u=1,\qquad a=1+\delta,\qquad |\delta|\ll1 .
\end{equation}
For $J<0$ the sixfold ground manifold splits by $2|\delta J|$. For $\delta>0$ the ground states are the four configurations in which the strengthened 1--3 bond is satisfied, and for $\delta<0$ they are the two configurations $(+-+)$ and $(-+-)$. In both cases the lowest excitation has the same magnetization content, $M=\pm1$, as the ground states.

\begin{figure}[t]
\centering
\includegraphics[width=0.9\textwidth]{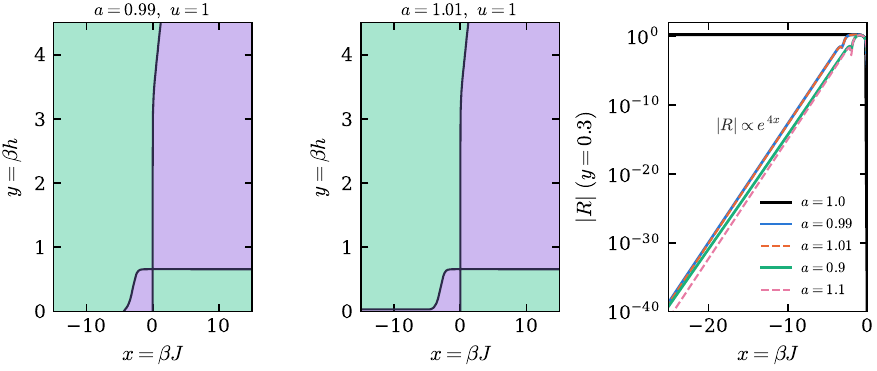}
\caption{Left and middle: sign of $R$ versus $x$ and $y$ for $u=1$ with $a=0.99$ and $a=1.01$. Right: $|R|$ versus $x$ at $y=0.3$ for $a=1$, $1\pm0.01$ and $1\pm0.1$.}
\label{fig:aniso}
\end{figure}

The sign maps for $a=0.99$ and $a=1.01$ are shown in the left and middle panels of figure~\ref{fig:aniso}, with the same colour code as in figure~\ref{fig:triangle}. They differ from the symmetric case only in the antiferromagnetic sector at $x\lesssim-2$. On the ferromagnetic side the line $y=y_c$ survives to high accuracy for $x\gtrsim1$, and the zero line near $x=0$ is only slightly bent. In the antiferromagnetic sector, by contrast, the zero line at $y_c$ disappears beyond $x\approx-4$. For $a=1.01$ a very thin negative strip survives at $y\lesssim0.05$, but there $|R|$ falls from $10^{-4}$ at $x=-4$ to $10^{-12}$ at $x=-8$. The sign maps alone thus exaggerate the difference between $a=0.99$ and $a=1.01$, since the curvature is numerically zero in most of the antiferromagnetic sector.

The right panel of figure~\ref{fig:aniso} makes this explicit on a logarithmic scale. For $a=1$ the curvature saturates at the frustrated plateau~(\ref{eq:Rminus}), $|R_{-\infty}(0.3)|=1.894$. For $a\neq1$ it follows the plateau only down to a crossover coupling $x^\ast$, beyond which it collapses as
\begin{equation}
|R|\propto e^{4x}=e^{-4\beta|J|}.
\end{equation}
The mechanism follows from equation~(\ref{eq:lowT}) and the remark after it. With $\delta\neq0$, $B$ fluctuates within the low-lying manifold, and these fluctuations soon dominate those caused by thermal excitation of the polarized states:
\begin{equation}
\langle\Delta B^2\rangle_{\rm split}\sim\delta^2,\qquad
\langle\Delta B^2\rangle_{\rm polarized}\sim e^{-4\beta|J|}.
\end{equation}
Because the split levels have identical magnetization content, they contribute to the metric but not to the leading curvature, and $R$ is left with the exponentially small contribution of the polarized states. The crossover therefore occurs where the two variances are comparable,
\begin{equation}
e^{4x^\ast}\sim\delta^2 .
\end{equation}
Locating $x^\ast$ numerically by the half-plateau criterion at $y=0.3$ for $10^{-6}\leq|\delta|\leq10^{-1}$, we find
\begin{equation}
x^\ast=\frac{1}{2}\ln|\delta|-0.24 ,
\label{eq:xstar}
\end{equation}
with the predicted slope $1/2$ to three significant digits and no dependence on the sign of $\delta$. In physical units this corresponds to the crossover temperature
\begin{equation}
k_BT^\ast\simeq\frac{2|J|}{\ln(1/|\delta|)} .
\label{eq:Tstar}
\end{equation}

Two conclusions follow. First, the limits $\delta\to0$ and $T\to0$ do not commute: the frustrated plateau is a nongeneric property of exact degeneracy. Second, since $T^\ast$ depends only logarithmically on $\delta$, the curvature detects a degeneracy lifting of relative size $\delta$ at temperatures far above the splitting itself, $k_BT^\ast\gg|\delta J|$. In this sense the curvature is an unusually sensitive probe of state selection in frustrated clusters.

\section{From the open chain to the triangle}\label{sec:chain}

\begin{figure}[t]
\centering
\includegraphics[width=0.85\textwidth]{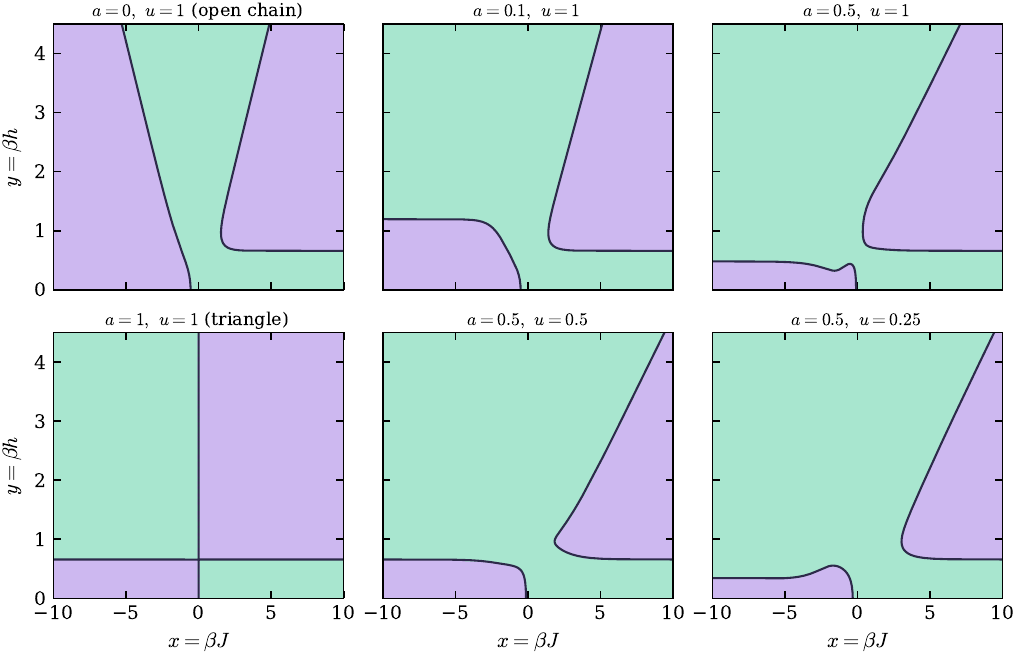}
\caption{Sign of $R$ versus $x$ and $y$. Top row: $u=1$ with $a=0$, $0.1$ and $0.5$. Bottom row: $(a,u)=(1,1)$, $(0.5,0.5)$ and $(0.5,0.25)$.}
\label{fig:chain}
\end{figure}

Figure~\ref{fig:chain} follows the sign structure along the path from the open chain to the triangle, $a=0$, $0.1$, $0.5$ and $1$ at $u=1$, and for two clusters with three unequal bonds, $(a,u)=(0.5,0.5)$ and $(0.5,0.25)$. The colour code is that of figure~\ref{fig:triangle}, and the solid lines mark the zero set of $R$. Three features are common to all clusters.

The first feature is a universal ferromagnetic limit. For $x\to+\infty$ the ground states are the polarized pair $M=\pm3$ for every $a,u>0$. The first excitation flips the spin with the smallest bond sum and therefore has $M=\pm1$. In equation~(\ref{eq:lowT}) this gives
\begin{equation}
\phi\propto\frac{\cosh y}{\cosh 3y},
\end{equation}
regardless of which spin flips and what it costs, so all clusters share the limit $R_{+\infty}(y)$ of equation~(\ref{eq:Rplus}). We verified this to four digits at $x=12$ for all couplings of figure~\ref{fig:chain}. The horizontal zero line at $y=y_c$ and the value $R=4/9$ at $h=0$ in the ferromagnetic sector are therefore universal.

The second feature concerns crossings of excited levels. In the anisotropic clusters a second zero line emerges from the ferromagnetic sector and becomes asymptotically straight,
\begin{equation}
y\simeq s\,x+\mathrm{const}.
\end{equation}
Its slope $s$ coincides with the ratio $h/J$ at which an $M=1$ one-spin-flip level crosses an $M=-1$ two-spin-flip level of the excitation spectrum. For the open chain, the excitation energies of the middle-spin flip and of the end-pair flip, measured from the polarized ground state, are
\begin{equation}
\Delta E_{\rm middle}=4J+2h,\qquad \Delta E_{\rm end\ pair}=2J+4h ,
\end{equation}
which cross at $h=J$; numerically we find $s=1.00$. For $a=0.1$ the corresponding crossing lies at $h=0.9J$ and we find $s=0.90$, while for $a=0.5$, $u=1$ it lies at $h=J/2$ and we find $s=0.50$. In the symmetric triangle all one-flip levels lie at $4J+2h$ and all two-flip levels at $4J+4h$. These never cross for $h>0$, consistent with the absence of a second zero line in the symmetric case. These zero lines are thus remnants of the zero-temperature level structure, pinned to fixed ratios $h/J$ rather than to fixed $y$.

The third feature concerns the antiferromagnetic sector. The open chain is not frustrated: its Néel ground states $(+-+)$ and $(-+-)$ satisfy both bonds. The first excitation, with one broken bond, again has $M=\pm1$, so the leading term~(\ref{eq:lowT}) vanishes and $R$ is exponentially small throughout the low-temperature antiferromagnetic region, as in section~\ref{sec:aniso}. On this side the zero line of the open chain follows
\begin{equation}
y\simeq|x|-0.80,
\end{equation}
that is, the field $h=|J|$ at which the one-broken-bond states become degenerate with the polarized state. Once $a>0$, the antiferromagnetic zero line instead becomes horizontal at low temperature, at the heights $y^\ast=1.195$ for $a=0.1$, $y^\ast=0.481$ for $a=0.5$ (both at $u=1$), $y^\ast=0.658$ for $(a,u)=(0.5,0.5)$ and $y^\ast=0.346$ for $(a,u)=(0.5,0.25)$. For $(a,u)=(0.5,0.5)$ the ground manifold has the same magnetization content as the frustrated triangle, with two states each of $M=\pm1$, and indeed $y^\ast=y_c$ within numerical accuracy.

Finally, the curvature at $x=0$ is nonzero for every anisotropic cluster; for the open chain, for example, $R(0,0.3)=0.478$, although the spins are free at $x=0$. This is no contradiction. On the manifold spanned by $(x,y)$ the coordinate $x$ is conjugate to the statistic $B$, and at $J=0$ the curvature reflects the third-order correlations of $B$ and $M$, which are nonzero for generic bond weights. The rule that $R=0$ for noninteracting systems applies to the geometry built on the noninteracting parameters themselves. It should not be applied pointwise on a manifold that also contains interacting directions.

\section{Discussion}\label{sec:discussion}

The three-spin cluster is small enough to be solved exactly, yet it contains the main ingredients for which thermodynamic geometry is usually invoked: competing interactions, a field that can reverse their effect, frustration and degeneracy lifting. It is therefore a useful benchmark for interpreting $R$ beyond the thermodynamic limit, and the exact results suggest several general lessons.

First, sign changes of $R$ in finite systems are generically simple zeros of an analytic function. Exponents extracted from equation~(\ref{eq:gamma}) are therefore equal to one and carry no information about criticality. What is informative is the location of the zero set and the magnitude of $R$. Here the zero set has a precise physical meaning: a single-spin cumulant condition, equation~(\ref{eq:kurt}), or the field-induced crossing of excited levels (section~\ref{sec:chain}). This parallels the role of the Widom line, along which curvature features persist beyond the critical point of fluids \cite{ruppeiner2012widom}, and the curvature-sign domains of finite Ising chains \cite{brody1999geometric}.

Second, sign maps should always be complemented by magnitudes. In large parts of the antiferromagnetic sector of anisotropic clusters $|R|<10^{-12}$, so sign changes there are physically insignificant and numerically delicate. Conversely, the frustrated triangle keeps $|R|$ of order one down to $T=0$. This plateau requires an exact degeneracy together with a lowest excitation that changes the magnetization content, equation~(\ref{eq:lowT}), and it is destroyed below a temperature that depends only logarithmically on the anisotropy, equation~(\ref{eq:Tstar}).

Third, the curvature depends on the chosen statistical manifold. Coordinates conjugate to different operators, here the distribution of $J$ over the three bonds, produce different geometries for the same physical point. The statement that $R$ vanishes for free spins holds only when the manifold is built from the noninteracting parameters.

These results bear directly on measurable systems. For a crystal of $N_{\rm mol}$ magnetically isolated trimers, the metric and the curvature per molecule are exactly those computed here, and all of them follow from second- and third-order cumulants, equation~(\ref{eq:Rcumulant}). These cumulants are accessible through magnetization, susceptibility and specific-heat measurements \cite{kahn1993,gatteschi2006}. The temperature-independent zero line at $\cosh 2y_c=2$ and the anisotropy-controlled collapse of the frustrated plateau, equation~(\ref{eq:Tstar}), are thus experimentally testable signatures, for instance in molecular triangles with slightly inequivalent exchange paths or in trapped-ion simulators of frustrated Ising triangles \cite{kim2010}. The analysis extends naturally to quantum spin trimers such as Heisenberg or XXZ triangles, where the noncommutativity of the Hamiltonian terms adds new geometric contributions, and to longer rings, in which the competition between the single-spin condition~(\ref{eq:yc}) and many-body correlations can be followed as the number of spins grows.

\section{Conclusions}\label{sec:conclusion}

We have determined the thermodynamic (Fisher--Rao) geometry of three Ising spins with arbitrary pairwise couplings in a field. The system is small, but its canonical description is exact. It describes macroscopic ensembles of independent spin trimers and engineered few-spin simulators, and it is the smallest cluster in which interaction topology and frustration can be varied continuously. For the symmetric triangle the scalar curvature has the closed form~(\ref{eq:Rtri}). Its zero set consists of the noninteracting line and the temperature-independent line $\cosh 2y_c=2$, the inflection point of the single-spin susceptibility, so the near-zero behaviour is exactly linear rather than critical. In the frustrated antiferromagnetic regime the curvature stays finite and negative down to zero temperature. A general low-temperature formula relates this plateau to the change in magnetization content between the degenerate frustrated manifold and its first excitation. Weak bond anisotropy destroys the plateau below the crossover temperature~(\ref{eq:Tstar}), after which $|R|$ decays as $e^{-4\beta|J|}$. Along the path from the open chain to the triangle, the ferromagnetic zero-temperature curvature is universal, whereas additional zero lines follow field-induced crossings of excited levels. These exact results provide a controlled reference for the use of thermodynamic curvature as a diagnostic of interactions, frustration and degeneracy in finite and small spin systems.

\section*{Declaration of competing interest}
The authors declare no competing interests.

\section*{Funding}
No funding was received for this work.

\appendix
\section{Derivation of the closed-form curvature}\label{app:closed}

We write
\begin{equation}
\ln\mathcal Z_\triangle=-x+\ln(2\cosh y)+\ln F,\qquad F=e^{4x}(2\cosh 2y-1)+3 .
\end{equation}
The linear term does not contribute to the metric. With
\begin{equation}
P=e^{4x}\cosh 3y,\qquad Q=3\cosh y,\qquad P+Q=F\cosh y,
\end{equation}
and
\begin{equation}
P_y=3e^{4x}\sinh 3y,\qquad Q_y=3\sinh y,
\end{equation}
the metric components are
\begin{equation}
g_{xx}=\frac{16PQ}{(P+Q)^2},\qquad
g_{xy}=\frac{4\left(P_yQ-PQ_y\right)}{(P+Q)^2},
\end{equation}
\begin{equation}
g_{yy}=\frac{(9P+Q)(P+Q)-(P_y+Q_y)^2}{(P+Q)^2},
\end{equation}
and the determinant is
\begin{equation}
G=\frac{144\,e^{4x}\left(3e^{4x}+2\cosh 2y-1\right)}{\cosh^2y\,\left[e^{4x}\left(2\cosh 2y-1\right)+3\right]^3}.
\end{equation}
Inserting these expressions into equation~(\ref{eq:R}) is simplest in the variables $W=e^{x}$ and $Y=e^{y}$, in which all entries are rational functions. The result is
\begin{equation}
R_\triangle=-\frac{(W^8-1)(Y^2+1)^2(Y^4-4Y^2+1)}{2\left(3W^4Y^2+Y^4-Y^2+1\right)^2},
\end{equation}
which is equation~(\ref{eq:Rtri}). We checked this expression against the cumulant form~(\ref{eq:Rcumulant}), evaluated by direct enumeration of the eight states, at randomly chosen points of the parameter plane.

\section{Low-temperature expansion}\label{app:lowT}

Let $|x|\to\infty$ at fixed $y$ and order the configurations by the value of $xB$. Denote by $B_0$ the extremal value of $B$, and suppose that the next value is separated from it by $\Delta>0$. Then
\begin{equation}
\ln\mathcal Z=xB_0+A(y)+\ln\left[1+\epsilon(x)\,\phi(y)+\cdots\right],\qquad \epsilon=e^{-|x|\Delta},
\end{equation}
with $A$ and $\phi$ defined in equation~(\ref{eq:Aphi}). To leading order in $\epsilon$,
\begin{equation}
g_{xx}=\Delta^2\epsilon\phi,\qquad g_{xy}=\mp\Delta\epsilon\phi',\qquad g_{yy}=A''+O(\epsilon),
\end{equation}
so that
\begin{equation}
G\simeq\Delta^2\epsilon\,\phi A'' .
\end{equation}
In the determinant of equation~(\ref{eq:R}), the terms of order $\epsilon^2$ proportional to $A'''$ cancel pairwise, and what remains is
\begin{equation}
\Delta^4\epsilon^2A''\left(\phi\phi''-\phi'^2\right).
\end{equation}
Dividing by $2G^2$ gives
\begin{equation}
R=-\frac{\phi\phi''-\phi'^2}{2\phi^2A''}+O(\epsilon)=-\frac{(\ln\phi)''}{2A''}+O(\epsilon),
\end{equation}
which is equation~(\ref{eq:lowT}). The result is independent of $\Delta$ and of an overall multiplicity factor in $\phi$. When $\phi$ is constant the leading term vanishes and $R$ is determined by higher excited manifolds, which explains the exponential suppression found in sections~\ref{sec:aniso} and~\ref{sec:chain}. For the ferromagnetic triangle,
\begin{equation}
A=\ln(2\cosh 3y),\qquad \phi\propto\frac{\cosh y}{\cosh 3y},
\end{equation}
which gives equation~(\ref{eq:Rplus}); for the frustrated triangle,
\begin{equation}
A=\ln(6\cosh y),\qquad \phi\propto\frac{\cosh 3y}{\cosh y},
\end{equation}
which gives equation~(\ref{eq:Rminus}).

\section*{References}
\bibliographystyle{iopart-num}
\bibliography{refs}

\end{document}